\documentclass[preprint,aps]{revtex4}

\usepackage{graphicx}
\usepackage{dcolumn}
\usepackage{bm}

\begin{document}


\title{Comment on ``On the classical analysis of spin-orbit coupling in hydrogenlike atoms,'' [Am. J. Phys. 78 (4) 428-432, April 2010]}

\author{David C. Lush } 
\affiliation{%
 d.lush@comcast.net \\}%


\date{\today}



\maketitle

In their recent paper \cite{KholmetskiiY10}, Kholmetskii, Missevitch, and Yarman ``reanalyze the usual classical derivation of spin-orbit coupling in hydrogenlike atoms'' and find a result ``in qualitative agreement with the solution of
the Dirac-Coulomb equation for hydrogenlike atoms.''  However, the authors' result is based on an equation of translational motion of the electron \cite{KholmetskiiY10_oc1} that omits any contribution due to the existence of ``hidden'' momentum of the electron intrinsic magnetic dipole moment in the electric field of the nucleus. As has been demonstrated \cite{ShockleyJames1967,Coleman1968,Hnizdo:1992}, accounting for hidden momentum is necessary to obtaining conservation of linear momentum in the interaction of a  classical current-loop magnetic dipole with a point charge. Current classical electrodynamics textbooks \cite{jcksn:classelec3,griffiths} also recognize this need. Additionally, it has been argued \cite{MunozY1} that hidden momentum of the electron intrinsic magnetic moment must be incorporated in the laboratory-frame analysis of atomic spin-orbit coupling, in order to obtain an equation of motion of the electron polarization that is consistent between the laboratory frame and the electron rest frame.  

Not taking issue with the authors' observation that the spin-orbit coupling magnitude must involve the magnitude of the binding force, including hidden momentum in the electron equation of translational motion has the effect of approximately halving  the non-Coulomb force on the electron compared to its value obtained omitting hidden momentum.  On the other hand, and apart from the issue of whether hidden momentum is associated with intrinsic as well as classical current-loop magnetic moments, if hidden momentum is omitted from the analysis, the force on the nucleus due to the electron will differ from the force on the electron due to the nucleus.   Thus, omitting the hidden momentum contribution, the binding energy including the spin-orbit coupling cannot be consistently calculated.  Furthermore, since the spin-orbit coupling magnitude calculated in Section III of the subject paper is based on the non-Coulomb force acting on the electron, it will be halved when hidden momentum is incorporated into the analysis.  The resulting spin-orbit coupling value will at that point be in disagreement with experiment.     

\section*{Acknowledgement}
\label{sec:Acknowledgement}

I wish to acknowledge the generous help of Drs. Kholmetskii, Missevitch, and Yarmin in correcting certain serious errors in this comment as originally submitted.

\bibliographystyle{plain}


\end{document}